\documentclass{aip-cp}

\usepackage[numbers]{natbib}
\usepackage{rotating}
\usepackage{graphicx}
\usepackage{bm}
\usepackage{url}

\begin{document}

% Title portion
\title{One-hundred-nm-scale electronic structure and transport calculations of organic polymers on the K computer}

\author[aff1]{Hiroto Imachi, Seiya Yokoyama, Takami Kaji, Yukiya Abe}
\author[aff2]{Tomofumi Tada}
\author[aff1]{Takeo Hoshi\corref{cor1}}

\affil[aff1]{Department of Applied Mathematics and Physics, Tottori University, Japan}
\affil[aff2]{Materials  Research  Center  for Element Strategy, Tokyo Institute of Technology, Japan}
\corresp[cor1]{Corresponding author: hoshi@damp.tottori-u.ac.jp}

\maketitle

\begin{abstract}
One-hundred-nm-scale electronic structure calculations were carried out on the K supercomputer by our original simulation code ELSES 
(\url{http://www.elses.jp/}) The present paper reports preliminary results of transport calculations for condensed organic polymers. Large-scale calculations are realized by novel massively parallel order-$N$ algorithms. The transport calculations were carried out as a theoretical extension for the quantum wavepacket dynamics simulation. The method was applied to a single polymer chain and condensed polymers. 
\end{abstract}

% Head 1==
\section{INTRODUCTION}

Organic materials play a crucial role among 
next-generation IoT products,
such as display, battery and sensor, 
since they form flexible atomic structures 
and enable ultra-thin, light, flexible (wearable) devices 
with a low fabrication cost. 
The atomic structure of organic device materials is disordered and
the simulation of such materials requires 
100-nanometer-scale systems,
which is  beyond the computational limit 
of the present standard electronic structure calculations.

Recently, electronic structure calculations with one-hundred-million atoms 
were realized on the K computer
by our electronic structure calculation code
ELSES (=Extra Large Scale Electronic Structure calculation; \url{http://www.elses.jp/}).
\cite{HOSHI-MARNOLDI, HOSHI-2013-JPSJ, HOSHI-2014-JPS-CP, HOSHI-2014-POS, IMACHI-2016-EIGENKERNEL}
The theory is based on generalized shifted linear equations
$((zS-H)\bm{x} = \bm{b})$, 
instead of the conventional generalized eigenvalue equation $(H \bm{y} = \lambda S\bm{y})$.
The computational cost is order-$N$ (${\rm O}(N)$) or proportional to the number of atoms $N$.
The method is extremely suitable to parallelism
and the benchmark on the K computer shows 
the efficient parallel computation in the strong scaling
with up to the full system (663,552 cores)
for one-hundred-million atoms (100nm-scale) materials.
Since the fundamental methodologies are purely mathematical,
they are applicable to various materials such as semiconductors and metals.
The code was developed for years,
by some of the authors (T. Hoshi and H. Imachi) and their co-workers,
particularly for academic-industrial collaboration researches.
A recent collaboration research is one for organic device materials
(Refs.~\cite{HOSHI-MARNOLDI, HOSHI-2014-JPS-CP}
and the present work) with Sumitomo Chemical Co., Ltd,
and another is one for battery materials 
\cite{NISHINO-ION-LIQUID, NISHINO-LISICON} 
with Toyota Motor Corporation. 
The other application researches can be found
in the reference lists of the above papers.
As technical details, 
the code uses {\it ab initio}-based model (tight-binding) theory
and, optionally, finer methodologies of
charge-self-consistent theory and van der Waals force.
A parallel direct solver the generalized eigenvalue equation 
was also implemented in our code.
\cite{IMACHI-2016-EIGENKERNEL}
The direct solvers are complement 
to the order-$N$ solvers,
because the direct solver gives numerically 
exact calculations with a heavier 
$({\rm O}(N^3))$  operation cost.

The present paper reports several preliminary results 
for the calculation of organic materials by ELSES
in quantum molecular dynamics simulations and transport (wavepacket dynamics) simulations.

\section{Quantum molecular dynamics simulation}

\begin{figure}[h]
  \centerline{\includegraphics[width=16cm]{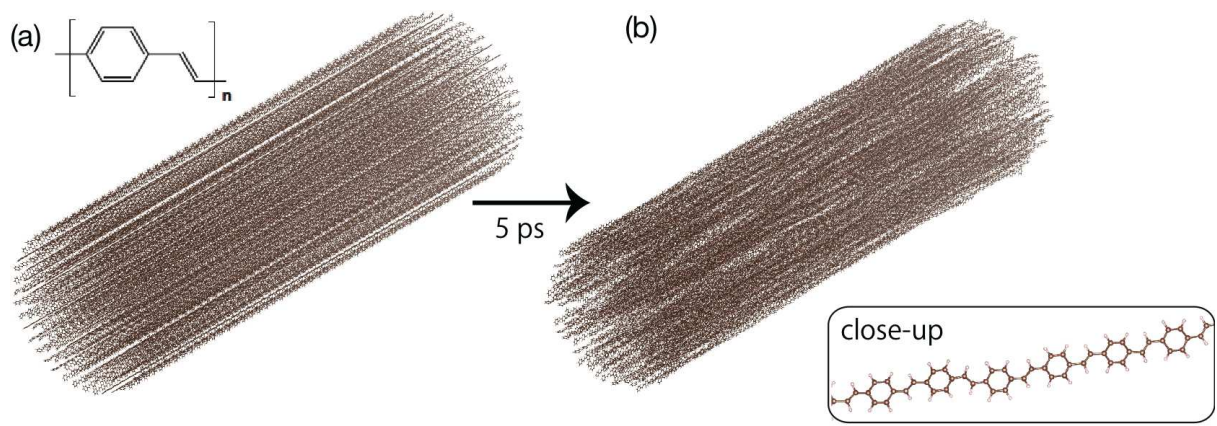}}
  \caption{Quantum molecular dynamics of bundle-like poly-(phenylene vinylene) (PPV)
 at the initial time ($t=0$) (a) and the final time ($t=5$ps) (b). 
  \label{FIG-QMD-PPV}
  }
\end{figure}

Quantum molecular dynamics simulations were carried out for condensed organic polymers.
The order-$N$ method was used.
See the previous paper \cite{HOSHI-MARNOLDI} for the methodological details. 
Figure  \ref{FIG-QMD-PPV} shows an example of bundle-like poly-(phenylene vinylene) (PPV).
The sample consists of 169 PPV polymers 
and each polymer has $n=50$ monomer units. 
The length of a polymer is $L \approx 40$nm.
The total number of atoms is $N$=117,962.
The van der Waals force \cite{ORTMANN} is included in the simulation. 
The parallel computation was carried out by 960 nodes ( 15,360 cores ) of 
the supercomputer Oakleaf-FX of the University of Tokyo.
The time interval of simulation step is $h_{\rm MD}=1$fs.
A finite-temperature simulation in $T$=600K was performed for 
$n_{\rm step}=5000$ iteration steps
or the period of $h_{\rm MD} \times n_{\rm step} =5$ps.
The total elapse time is approximately 10 hours.

\section{Transport calculation with wavepacket dynamics}

\begin{figure}[h]
  \centerline{\includegraphics[width=16cm]{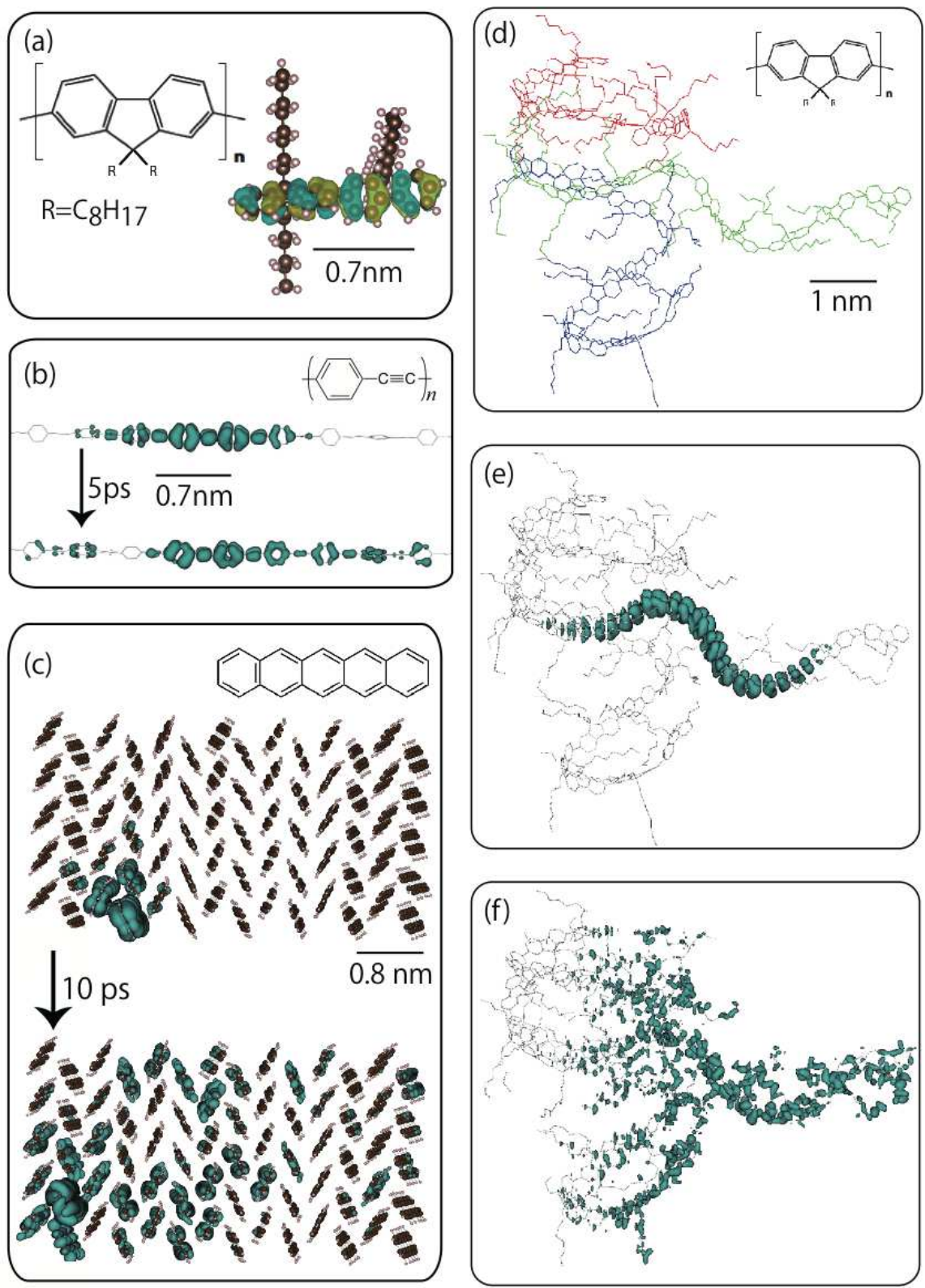}}
  \caption{(a) 
   Example of organic polymer: poly-((9,9) dioctyl-fluorene) (PFO) with $n=2$.
   The HOMO state is drawn as a $\pi$ wavefunction. \cite{HOSHI-2014-POS} 
  (b) Quantum wavepacket dynamics on a polymer of poly-(phenylene–ethynylene).  The figre is a close-up . (c) Quantum wavepacket dynamics on pentacene thin film with single layer. (d) A condensed polymer of poly-((9,9)-dioctyl fluorene)(PPE). The three polymers in a periodic cell form an amourphous-like strcutures. The three polymers are painted in different colors for better understanding. (e)-(f) Quantum wavepacket dynamics on the condensed polymers. The initial state at $t=0$ (e) and the final state at $t=1$ps (f) are drawn. 
  \label{FIG-WPD}
  }
\end{figure}

Quantum wavepacket (WP) dynamics simulations
were also carried out for the investigation of transport property
on the K computer and the other supercomputers. 
The transport of organic devices is governed
by $\pi$ wavefunctions. 
Figure \ref{FIG-WPD}(a) shows 
a $\pi$ wavefunctions in an organic polymer,
poly-((9,9)-dioctyl fluorene) with $n=2$. 
Both ballistic and non-ballistic conduction mechanisms are important.
In a fairly disordered sample, for example, 
$\pi$ wavefunctions are so localized that they are hard to propagate
and the sample will show a smaller value of mobility.
The simulation is based on a Schr\"odinger-type equation 
($\partial_t \Psi = - i H_{\rm WP} \Psi$)
for a hole wavepacket $\Psi(\bm{r},t)$
with a modelled Hamiltonian $H_{\rm WP}$. 
The time interval of simulation step  is 
typically, $h_{\rm WP}=0.1$fs and
is much smaller than that in the molecular dynamics simulation
 ($h_{\rm WP} \ll h_{\rm MD}$).
In a previous paper, \cite{TERAO-2013}
a WP dynamics simulation was carried out
for organic polymers,
in which a $\pi$-orbital-only model is constructed 
for disordered atomic structures in dynamical simulations.
The analysis of the result gives mobility.  
The present method is a theoretical generalization of the previous one \cite{TERAO-2013}.
Methodological details have not yet settled and a brief outline is explained here.
The initial wavepacket $\Psi(\bm{r},t=0)$ for a hole 
is set as an eigenstate, such as the HOMO state.
Now we are constructing  
two kinds of dynamical simulation methods;
The one is 
the combined simulation between molecular dynamics and wavepacket dynamics simulations,
or the explicit dynamical simulation of atoms and the wavepacket. 
The other is the wavepacket dynamics simulation 
in which the atomic motion is treated as a modeled perturbed term 
in the wavepacket dynamics.
The latter method gives a faster simulation.
Since our purpose is the systematic survey among  different samples
the problem is the balance between the computational cost and the accuracy.
If the simulation runs faster, 
we can investigate a dynamics in a larger time scale
or many sample in different structures and/or conditions. 

Several preliminary results are discussed below.
The wavepacket dynamics simulation was carried out
for organic polymer of poly-(phenylene–ethynylene) (PPE).
Single polymer simulations were carried out
among different conditions and polymer lengths.
The maximum length of polymer is $L \approx$ 700nm
with $n=1,000$ monomer units.
An example is shown in Fig.~\ref{FIG-WPD}(b) as a close-up. 
A characteristic semi-localized $\pi$  wavepacket was observed,
since the wavepacket spreads over several monomer units
butnot over the whole region of the polymer.
Our preliminary results on calculated mobility 
are consistent to the experimental trend \cite{TERAO-2013}
in which the mobility of meta(zig-zag) type polymers is larger 
than that of para (linear) type polymers. 
Figure \ref{FIG-WPD}(c) shows the wavepacket dynamics simulation
in a disordered pentacene thin film with a single layer.
The periodic boundary condition is imposed
on in the upper and lower area in Fig.~\ref{FIG-WPD}(c).
The $\pi$ wavefunction propagates 
from a molecule into another. 
Figure \ref{FIG-WPD}(d)-(f) 
shows condensed polymers
of poly-((9,9)-dioctyl fluorene). 
%The initial structure was generated
%by a classical molecular dynamics simulation.
Three polymers in a periodic cell 
form an amourphous-like structure, as shown in Fig. \ref{FIG-WPD}(d).
The number of atoms in a polymer is $N_{\rm polymer}=692$
and the total number of atoms is $N = 692 \times 3 = 2076$.
Figure \ref{FIG-WPD}(e)-(f) shows 
a result of wavepacket dynamics.
The initial wavepacket is localized in one polymer
and the final one extends over polymers.
More quantitative discussions are under development.

%%%%%%%%%%%%%%%%%%%%%%%%%%%%%%%%%%%%%%%%%%
% Figure

\section{Summary and future outlook}

Novel linear algebraic algorithm realizes 
100-nm-scale electronic structure calculations.
The code was applied to 
organic device materials
in quantum molecular dynamics and quantum wavepacket dynamics simulations.
As a future outlook,
the convergence between simulation and data sciences will be realized. 
The present simulation code can generate huge data of electronic wavefunctions 
in disordered 100-nm-scale materials. 
We should analyze the materials systematically
and find an insight for better device material and/or its fabrication process.

% Sections that will go in second font

% Acknowledgement
\section{ACKNOWLEDGMENTS}
The authors thanks to Masaya Ishida (Sumitomo Chemical Co.),
since he provided us several atomic structure data. 
This research is partially supported also by Grant-in-Aid 
for Scientific Research
(KAKENHI Nos. 26400318 and 26286087)
from the Ministry of Education, Culture, Sports, Science and Technology 
(MEXT) of Japan. 
The K computer was used in the research project numbers of hp150144 and hp150281. 
The supercomputer Oakleaf-FX of the University of Tokyo was used 
in Initiative on Promotion of Supercomputing for Young or Women Researchers,
Supercomputing Division, Information Technology Center, The University of Tokyo.
We also used the supercomputer 
at the Institute for Solid State Physics of the University of Tokyo and 
at the Research Center for Computational Science, Okazaki.

% References

%\nocite{*}
%\bibliographystyle{aipnum-cp}%
%\bibliography{sample}%

\end{document}